Privatdozent Dr. Harry Mönig
Physikalisches Institut
Universität Münster
Wilhelm Klemm Straße 10, D-48149 Münster
harry.moenig@uni-muenster.de
Homepage

**Harry Mönig**

# IMAGING ATOMS IN REAL-SPACE WITH ELEMENTAL SELECTIVITY

In scanning probe microscopy experiments, the identity of the tip-terminating atom and its bonding configuration at the apex can drastically affect the image contrast [1-3]. In particular, in the field of noncontact atomic force microscopy (nc-AFM), considerable progress has been made through the implementation of different tip functionalization approaches, which, in combination with the ultra-small oscillation amplitudes of qPlus-type force sensors, have led to a new era in surface chemistry [4,5]. By far the most established probes used in nc-AFM experiments are CO-functionalized tips. They are prepared by first co-adsorbing a low density of CO molecules on the surface under study and then picking up a single one by a sharp metallic tip apex leading to a configuration where the O-atom points to the surface [1,4,5]. This significantly reduces chemical interactions within the tip-sample junction allowing for constant-height nc-AFM imaging in the repulsive force regime where organic nano-structures can be imaged with sub-molecular resolution [4,5]. However, a major problem of using such probe particles for nc-AFM experiments is their lateral flexibility at the apex. It can lead to image distortions, a systematic overestimation of bond lengths, and even artificial contrast features [6-8]. The latter particularly concern surface sites with a strongly varying tip-sample potential where tip flexibility affects the contrast the most.

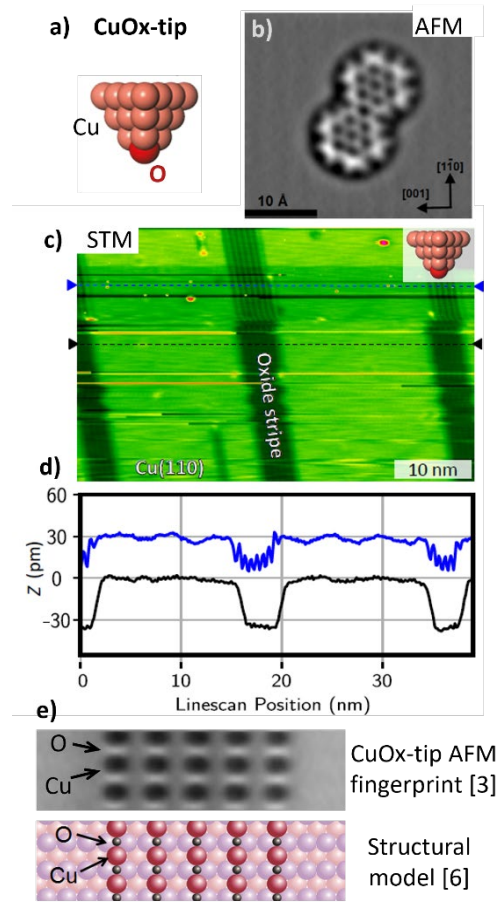

In the following, an alternative approach based on so called copper-oxide tips (CuOx-tips) is discussed. These tips consist of a Cu-apex terminated by a single O-atom, which is covalently bound in a tetrahedral configuration to its upper Cu-atoms (Fig. 1a). With this structure, CuOx-tips allow for stable constant-height imaging in the repulsive force regime and to resolve single bonds in organic nano-structures (Fig. 1b). At the same time, CuOx-tips are extremely rigid, which allows to largely neglect artefacts due to tip flexibility [3,6,7].

CuOx-tips can be prepared by controlled tip indentations and voltage pulses on a slightly oxidized single crystalline Cu substrate. Figure 1c shows a scanning tunneling microscopy (STM) image of a partially oxidized Cu(110) surface showing the typical (2x1)O-reconstructed oxide stripes, which is acquired during such tip forming procedures [9]. While recording the lower part of the image, the tip is terminated in a pure metallic apex where frequent tip indentations and voltage pulses below the scan line result in instabilities. Due to the reduced density of states, the topographically elevated oxide stripes appear as deep trenches in the STM contrast at feedback settings of -0.1 V/50 pA (see also black line profile in Fig. 1d). Only when the tip picks up copper-oxide material from the surface leading to an O termination at the apex, the STM contrast changes significantly (upper part of Figure 1d). This not only increases the resolution, but also reduces the apparent depth of the trenches (blue line profile in Fig. 1d). Yet, it is important to note that such a change in the STM contrast alone does not prove a CuOx-tip termination. In principle, any passivated tip could show such a contrast. To unambiguously verify a CuOx-tip termination the recording of an nc-AFM image above one of the oxide stripes is required. As demonstrated by systematic DFT simulations, only a contrast as shown in Fig. 1e allows to confirm the formation of a CuOx-tip. In fact, a constant-height nc-AFM contrast as shown in Fig. 1e can be used as fingerprint for a CuOx-tip [3,6,7]. Here, single Cu-atoms within the (2x1)O-reconstruction appear as dark ellipsoidal depressions, while the O-atoms are imaged as bright contrast features (dominated by repulsive forces) in between (Fig. 1d). As shown in the following, this remarkable chemical selectivity for metal- and oxygen atoms is a specific property of CuOx-tips, which can even be generalized for the whole class of metal-oxide surfaces.

As a first step for this generalization, the role of the tip termination was investigated by comparing the contrast of four commonly used probe tips. The upper panels in Fig. 2 show constant-height nc-AFM images of (2x1)O-reconstructed oxide stripes recorded with four atomically defined tip-terminations at tip-heights at which the maximum contrast could be achieved in each case [3]. It was found that the high reactivity of pure Cu-tips (Fig. 2a) leads to atomic relaxations within the tip-sample junction already in the attractive force regime. Therefore, imaging at even more reduced tip-sample distances leads to unstable imaging conditions and sudden tip changes. For the case of Xe- and CO-passivated tips, it is possible to enter the repulsive force regime (Fig. 2b,c). However, the observed contrast does not allow to clearly identify specific atomic surface sites, rather the images appear strongly distorted and show emphasized contrast signatures at inter-atomic sites (i.e. between the O-Cu-O rows). Image simulations allowed to explain these findings by dynamic tip bending effects, which are particularly emphasized due to the strongly varying tip-sample potential between Cu- and O-sites on the surface during scanning [3]. On the contrary, the data recorded with a CuOx-tip show a pronounced site-specific interaction and shape distinction for the Cu- and O atoms within the (2x1)O-reconstructed oxide stripes (Fig. 2d). Therefore, the identification of Cu- and O sites is straight forward. As demonstrated by height-dependent imaging in combination with force-distance spectroscopy, the observed elemental selectivity is obtained over an extended range of tip-heights [3].

Fig. 1: a) DFT-optimized model of a CuOx-tip. b) Constant-height nc-AFM image of dicoronylene on Cu(110) recorded with a CuOx-tip (reprinted from ref. [6] with permission from Springer Nature). c) Constant-current STM image of a partially oxidized Cu(110) surface recorded during tip forming procedures [9]. d) Line profiles from c for a pure metallic apex (black) and after successful CuOx-tip functionalization (blue). e) Constant-height nc-AFM on an oxide stripe (adapted from ref. [3] with permission from Royal Society of Chemistry). Such a contrast serves as a fingerprint allowing to unambiguously verifying a CuOx-tip. Structural model reprinted from ref. [6] with permission from Springer Nature.

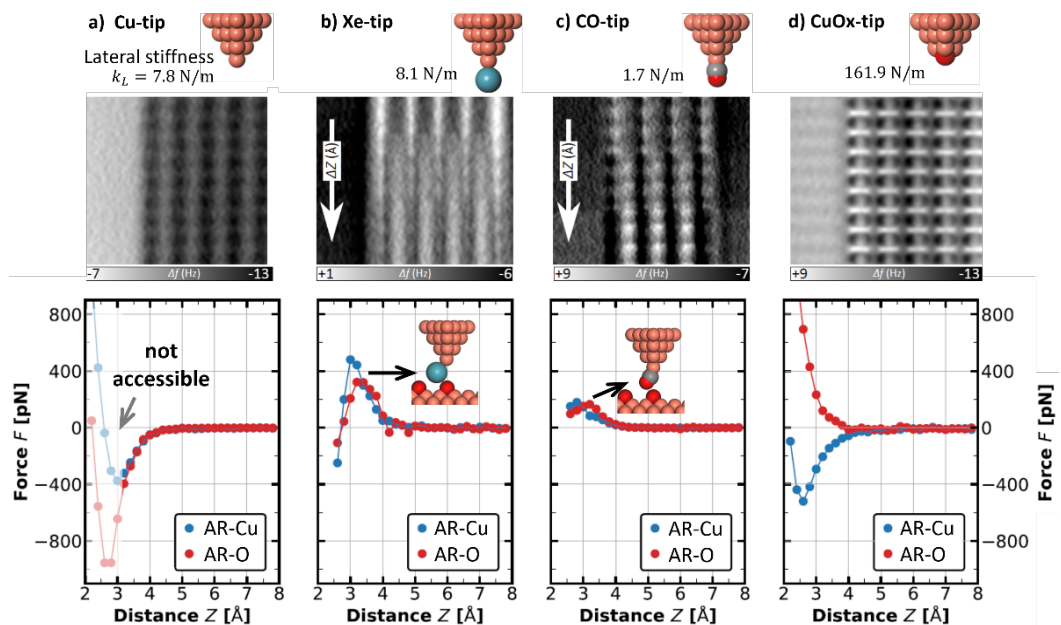

Fig. 2: (Upper panels) Comparison of nc-AFM images for various atomically defined tip terminations, recorded in constant-height mode above (2x1)O reconstructed oxide stripes of a partially oxidized Cu(110) surface. The DFT-derived lateral spring constants given next to the tip models allow to quantify tip flexibility. (Lower panels) DFT-simulated F(Z) force-distance spectra where the vdW components have been subtracted to assess the chemical interactions for O- and Cu sites. Adapted from ref. [3] with permission from Royal Society of Chemistry.

Force-distance (F(Z)) simulations based on density functional theory (DFT) provide a more detailed view on the contrast mechanism. Specifically, the lower panels in Fig. 2 show corresponding F(Z) curves from O- and Cu-sites for all the four tips. To directly compare the chemical interactions, the simulated data are plotted without van der Waals contributions and on identical force scales. On both atomic sites, the Cu-tip shows pronounced minima at Z < 3 Å, which can be associated with a strong attraction and related high chemical reactivity, especially on the O-site. With the differing force interaction of the Cu- and O sites, such force curves indicate a strong contrast at Z < 3 Å, which in principle would allow to clearly distinguish the atoms within this force regime. However, the high chemical forces in combination with a weak stiffness of the Cu-tip leads to atomic relaxations within the tip-sample junction. Therefore, the tip-height regime in which a strong contrast could be expected in principle is not accessible without the metallic tip reacting with surface oxygen [3,7]. On the contrary, the Xe- and CO-tip both hardly show any attraction, which confirms the chemical passivation of these two tips. At the same time, the force-distance curves show nearly no separation, which explains the experimentally observed weak contrast between the O- and Cu-sites. In addition, both show kink-like maxima around Z = 3 Å, which are typical for lateral bending effects of the Xe- and CO probe particles [3,8]. In comparison, the curves for the CuOx-tip show a largely inert interaction on the O-site and a moderate attraction (i.e. reactivity) on the Cu-site, leading to a distinct separation of the F(z) curves (Fig. 2d), which is in agreement with the experimentally observed strong contrast. Furthermore, due to the high stiffness of the CuOx-tips, artefacts due to tip bending effects are absent. The differences in tip flexibility are demonstrated by a quantitative comparison of the DFT-derived lateral spring constants, given in the top row of Fig. 2 [3].

Up to this point, elemental selectivity was observed only for the (2x1)O-reconstructed Cu(110) surface. Consequently, our next steps addressed the question if this finding also applies to other reconstructions and other metal oxide surfaces. To systematically investigate this, CuOx-tips were applied to various O-induced reconstructions, namely Cu(110) (6x2)O, Cu(100) (2$\sqrt{2}$x$\sqrt{2}$)R45°O, and Ag(111) p(4x4)O [10]. Even for these surfaces where Cu- and O atoms occupy sites with a variety of different relative heights, the distinct elemental selectivity of the contrast allowed to directly identify atomic sites and defect structures. Based on these results, height-dependent image data were compared with corresponding DFT-derived maps of the local electrostatic potential. The excellent agreement of theoretical and experimental data, exemplarily shown in Fig. 3 a and b for the Cu(100) (2$\sqrt{2}$x$\sqrt{2}$)R45°O surface and an OH defect on (magnetite) $Fe_3O_4$(001), allowed to conclude that the elemental selectivity is dominated by electrostatic interactions, which strongly correlates with the surface charge density [10]. In particular, this concerns the interactions between the negative charge of the tip-terminating O-atom and the strongly varying electrostatic potential of metal- and oxygen sites of the surface (Fig. 3c).

Extending this methodology to surfaces of bulk metal oxide materials and highly complex surface structures, demonstrates the powerful capability of this approach. Figure 3d shows data of a titanium oxide (TiOx) thin film grown on a Pt(111) substrate where the arrangement of 4-fold and 3-fold coordinated titanium atoms leads to a surface structure with a high degree of complexity [10]. Similarly, Figure 3e shows CuOx-tip measurements of an $Al_2O_3$(0001) surface which forms a ($\sqrt{31}$x$\sqrt{31}$)R±9° reconstruction exhibiting a stochiometric composition and emphasizing the applicability also to insulating surfaces [12]. In these experiments, the distinct contrast obtained with the CuOx-tip provided immediate access to the metal- and oxygen sublattices. Beyond that, imaging with such an elemental selectivity provides access to a direct structural characterization of atomic-scale point defects (Fig. 3b) [10,12], which play an important role in the fundamental understanding of the catalytic, optical, and electronic properties of metal-oxide materials [10,11,13].

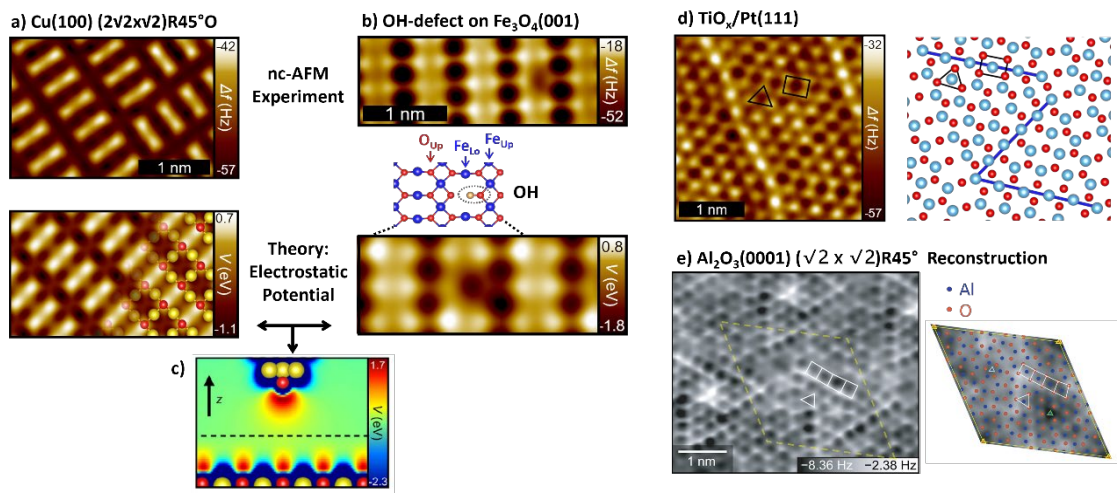

Fig. 3: a) and b) Comparison of experimental CuOx-tip nc-AFM data with the calculated electrostatic potential for DFT optimized surface models. c) Cross section of the electrostatic potential of the CuOx-tip and a metal oxide surface. d) nc-AFM data of a titanium oxide thin film grown on Pt(111). The elemental selectivity of the CuOx-tip allows to directly obtain the surface structure (shown on the right). e) CuOx-tip nc-AFM image recorded on the bulk insulator $Al_2O_3$ (left). The surface structure can be directly derived from the experiment (right). a-d: Reprinted with permission from ref. [10]. Copyright 2014 American Chemical Society. e: From ref. [12], reprinted with permission from AAAS.

With its universal elemental selectivity on metal-oxide surfaces, CuOx-tip functionalization constitutes a powerful methodology for the atomic-scale characterization for this technologically relevant class of materials. The successful application to even highly complex surfaces and defects where conclusive structural models had been missing so far, must be set in context to the longstanding dream for scanning probe microscopy to directly determine the elemental identity of single atoms [14-16]. Our approach allows to drastically reduce the theoretical efforts and related indirect structural considerations and assumptions (e.g., for tip terminations and atomic positions) for the development of structural surface models. Moreover, CuOx-tips allow neglecting imaging artefacts due to tip-flexibility [6,7] and can be routinely used at liquid nitrogen temperatures as for example the experimental data in Fig. 3 a-d [10,17]. These highly desired properties make these probes unique for standardized scanning probe microscopy experiments.